\documentclass[namedreferences]{solarphysics}
\usepackage[hyperref,optionalrh,solaromanenum]{spr-sola-addons} % For Solar Physics 
\usepackage{graphicx}                    % For eps figures, newer & more powerfull
\usepackage{color}                       % For color text: \color command
\begin{document}

\begin{article}

\begin{opening}

\title{Origin of a CME-related shock within the LASCO C3 field-of-view}

%%%%%%%%%%%%%%%%%%%%%%%%%%%%%%%%%%%%%%%%%%%%%%%%%%%
%% Authors Names
%
\author[addressref={iszf},corref,email={vfain@iszf.irk.ru}]{\inits{}\fnm{V.G.}\lnm{Fainshtein}}
\author[addressref={iszf},corref,email={egorov@iszf.irk.ru}]{\inits{}\fnm{Ya.I.}\lnm{Egorov}}

%%%%%%%%%%%%%%%%%%%%%%%%%%%%%%%%%%%%%%%%%%%%%%%%%%%
%% Runningheads
%
\runningauthor{Fainshtein et al.}
%\runningtitle{Fainshtein et al., 2015}

%%%%%%%%%%%%%%%%%%%%%%%%%%%%%%%%%%%%%%%%%%%%%%%%%%%
%% Affilations 
%% id shold be the same with \author addressref value.
\address[id=iszf]{Institute of Solar-Terrestrial Physics SB RAS, PO Box 291, Irkutsk, Russia}
%\address[id=egorov]{Institute of Solar-Terrestrial Physics SB RAS, PO Box 291, Irkutsk, Russia}
%\address[id=rud]{Institute of Solar-Terrestrial Physics SB RAS, PO Box 291, Irkutsk, Russia}
%\address[id=anf]{Institute of Solar-Terrestrial Physics SB RAS, PO Box 291, Irkutsk, Russia}

%%%%%%%%%%%%%%%%%%%%%%%%%%%%%%%%%%%%%%%%%%%%%%%%%%%
%%% Abstract 
\begin{abstract}
We study the origin of a CME-related shock within the LASCO C3 field-of-view (FOV). A shock originates, when a CME body velocity on its axis surpasses the total velocity $V_A + V_{SW}$, where $V_A$ is the Alfv\'en velocity, $V_{SW}$ is the slow solar wind velocity. The formed shock appears collisionless, because its front width is manifold less, than the free path of coronal plasma charged particles. The Alfv\'en velocity dependence on the distance was found by using characteristic values of the magnetic induction radial component and of the proton concentration in the Earth orbit, and by using the known regularities of the variations in these solar wind characteristics with distance. A peculiarity of the analyzed CME is its formation at a relatively large height, and the CME body slow acceleration with distance. We arrived at a conclusion that the formed shock is a bow one relative to the CME body moving at a super Alfv\'en velocity. At the same time, the shock formation involves a steeping of the front edge of the coronal plasma disturbed region ahead of the CME body, which is characteristic of a piston shock.
\end{abstract}

%%%%%%%%%%%%%%%%%%%%%%%%%%%%%%%%%%%%%%%%%%%%%%%%%%%
%% Keywords
%

\end{opening}
%-------------------------------------------------

%%%%%%%%%%%%%%%%%%%%%%%%%%%%%%%%%%%%%%%%%%%%%%%%%%%
%% Sections
%
\section{Introduction}
     \label{S-Introduction} 

The assumption that, ahead of solar material bunches that were earlier thought to be emitted during solar flares, a shock may emerge in the interplanetary space was already stated in 1950s \citep{Gold1962}. Based on observations of coronal mass ejections (CMEs) in 1973-1974 (Skylab era),  \citealp{gosling1976} supposed that a bow shock should emerge ahead of sufficiently fast CМEs. Later, the assumption of a shock emergence ahead of fast CМEs was stated repeatedly. That assumption leaned on different data, but only indirect or proxy evidence was provided. \citealp{Sheeley2000} detected that, ahead of fast CMEs, there are disturbances, propagating away from such mass ejections in white light coronal images from the SOHO Large-Angle Spectrometric Coronagraph (LASCO; \citealp{lasco}). Such disturbances were assumed to be shocks in case, when the disturbance velocity surpassed the fast-mode MHD velocity. Earlier, \citep{Hundhausen1987} assumed that, as long as the CME velocity exceeds the speed of sound or the Alfv\'en velocity, such a CME is shock-related. Also, coronal structures observed at CME motion \citep{Sheeley2000} and, finally, recorded type-II radio bursts, whose sources CME-related shocks are considered to be, may be indirect proof of shock existence (see reviews by Gopalswamy, 2011; 2016, and references therein).

For the first time, a CME-related shock (in the form of the electron density jump) was detected on the images of the white corona within the LASCO field-of-view \citep{Vourlidas2003}. Today, shocks are observed ahead of fast CMEs within the LASCO C2 and/or C3, as well as other coronagraphs, fields-of-view \citep{Vourlidas2009,Vourlidas2012}. Such shocks are referred to as CME-driven shocks. A weak brightness front recorded ahead of the CME bright frontal structure boundary was noted to be a morphological feature observed on white corona images involving CMEs. Also, the space between the boundary and the front features a weak emission \citep{Manchester2008}. And, finally, in \citep{Ontiveros2009}, for several CMEs, electron density distributions along the radial direction were obtained from the brightness distributions in the region of a weak external front ahead of a CME. On these electron density distributions, a density jump that is considered as a shock was distinctly identified. Note that the spatial width of such density jumps is comparable with the LASCO C3 spatial resolution.

Known are, at least, four physical mechanisms for shock generation. Shocks may be generated by an impermeable piston (plane, cylindrical, spherical, etc.) compressing (as it moves) the surrounding medium, and the subsequent evolution of the originating disturbance leads to a shock production (piston shocks) \citep{Sedov,Zeldovich}. A shock may be a bow one originating as the surrounding gas (plasma) flows around a body at supersonic velocity \citep{Landau_EN}. A shock may be explosive originating due to an abrupt change in the macroscopic parameters of the medium (temperature, pressure, etc.) in a limited region, and due to a subsequent expansion of the latter \citep{Sedov,Zeldovich}.

Several years ago, another view of the CME-related shock nature was stated. Such shocks (at least, some of them) are generated by fast solar filaments preceding CME emergence. These filaments move at a large acceleration. In general case, shocks are generated by flux-ropes in the lower corona that disturb the latter \citep{Grechnev2011,Grechnev2016}. At the initial stage of their existence, such chocks behave as decelerated blast waves. Particularly, in \citep{Fainshtein2015}, for the 2010 June 13 event, the time dependence of the shock position and of the shock velocity obtained from the SDO AIA was established to agree with theoretical time dependences of these motion characteristics within the explosive shock self-similar motion. As it moves away from the Sun, within the LASCO C2 or C3 FOV, such a shock may appear related to a CME, and further behave like a CME-driven shock.

Until now, there have been practically no investigations to reveal the nature of individual shocks recorded within the LASCO C2, C3, etc., FOV. \citealp{Vourlidas2003} arrived at a conclusion that the shocks recorded ahead of CMEs are bow shocks. In the review by \citep{Chen2011}, such shocks are referred to as piston. In fact, CME-related shocks may exhibit signatures of both bow shocks (because the surrounding plasma flows around of the CME body moving translatorily), and piston shocks (because the CME body expands simultaneously with its translational motion, which leads to the surrounding plasma compression). Hereinafter, we refer to the CME part restricted by the frontal structure outer boundary as a CME body. One can establish the physical mechanism for a CME-related shock generation, as long as one studies details of a bow shock formation. As far as we know, there have been no such investigations until recently.

Finally, we note that, currently, by using the data from different instruments, it is possible to determine both geometrical (3D shock position and shape) and physical (density and temperature jumps, as well as the magnetic field at the shock front) shock characteristics \citep{Mancuso2009,Ontiveros2009,bemporad2010,Bemporad2014,Susino2015,Fainshtein2015}.

In this study, we investigate the origin of a CME-related shock within the LASCO C3 FOV, and reveal, to which of the four above mechanisms this shock formation may be related to. For that purpose, we analyze the shock formation process. Also, revealed is whether the coronal plasma particle collisions determine the width of the originated shock front, or the shock is collisionless.

\section{Data Analysis Methods}
     \label{S-Data} 
In order to find the moment of CME-driven shock origination the CME have to be fast enough to generate such a shock and slow enough to be captured by LASCO in several moments of time.
For the analysis, we selected the CME originated in AR NOAA 11521, and first recorded within the LASCO C2 FOV on 2012 July 17 (13:48:06 UT). This mass ejection originated relative high in the lower corona, its velocity grew slowly with time/distance at the initial stage of its motion. This CME appeared to be related to the X-ray flare, whose class is difficult to estimate. The emission maximum within the (1-8) \AA wavelength range (the GOES flare class was М2.0 at that moment) appeared at about 17:20 UT, when the CME had already left the LASCO C2 FOV. Herewith, the change in the X-ray intensity, Isxr(t), occurred non-monotonously, with several maxima and inflections at the Isxr front. Apparently, the recorded SXR flux was related to two CMEs originated at close instants (both CMEs are indicated at \url{https://cdaw.gsfc.nasa.gov/CME_list/}).

\begin{figure}[!ht] 
\centerline{\includegraphics[trim=0.0cm 0cm 0.5cm 14cm, width=1.0\textwidth]{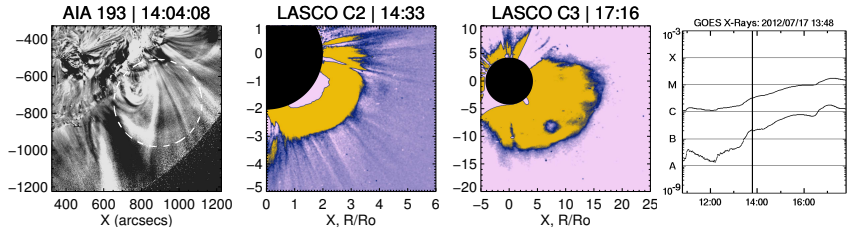}}
\caption{a) the CME during eruption from observations in the 193\AA~ channel; b,c) CME observed with the LASCO C2 and C3 coronagraph. d) the soft X-ray flux within the (1-8)\AA~ wavelength range (GOES);}%\label{fig:?}
\end{figure} 

To investigate the CME-related shock origin and evolution, we used calibrated white light coronal images obtained with the LASCO С3, at the Level 1 processing (\url{https://sharpp.nrl.navy.mil/cgi-bin/swdbi/lasco/images/form}). From these images, we formed the relations of the corona images, when the images containing the CME were split in each pixel into the corona image before the CME emergence within coronagraph FOV. 

To detect a shock, we used another technique based on subtracting the background determined from averaging several tens of images within 2.5 hours prior to the flare onset from the source images of the corona. The image points with value less than standard deviation were set to zero. Using no other averaging in the surroundings of each dot on the brightness scan line further, we determined through this approach the minimal observed width of the brightness jump that was identified as a shock (see Section 3).

While revealing the physical mechanism for the CME-related shock formation, we endeavored to answer the following question: Does a CME-related shock originate at the distance, where the $V_b > V_{MS} + V_{SW}$ condition is met? Here, $V_b$ is the CME body velocity, $V_{MS}$ is the speed of fast magnetic sound, $V_{SW}$ is the velocity of the surrounding solar wind. At the distance that we are interested in $(5-25)R_s$, where $R_s$ is the Sun radius, $V_{MS} \approx V_A$, $V_A$ being the Alfv\'en velocity.

To estimate the location, where the $V_b > V_{MS} + V_{SW}$ condition is met, one should know the R-distance ($R$ is counted off from the solar disk center) dependences of the Alfv\'en velocity and of the velocity of the surrounding solar wind. Sometimes, to estimate the $V_A(R)$ dependence, used is the $V_A$ dependence on the distance obtained in \citep{Mann1999} for quiet regions of the solar atmosphere (see, for example, \citealp{Kim2012}). The authors themselves in \citep{Mann1999} referred to the obtained $V_A(R)$ dependence as a crude estimate due to a strong non-uniformity of the corona. Thus, for example, the crude B estimate at the photosphere $(B=2.8 G)$ in that paper (that, indeed, may significantly vary from $2.8 G$ at different sites of the quiet corona) affects the $V_A(R)$ dependence from \citep{Mann1999}. Besides, the $V_A(R)$ dependence obtained by \citealp{Mann1999} is based on the authors' incorrect notions about the behavior of the quiet corona magnetic field. The authors believe that $B \approx 1/R^2$ at all the heights over the photosphere in the quiet corona. In fact, there is no such a distance dependence of the magnetic field, at least, at the heights from the photosphere to the $R=(2.5 - 3)R_s$ distance. Indeed, the $B \approx 1/R^2$ dependence is possible only in the region of the open field lines directed radially. In this case, the magnetic flux in any magnetic tube in this region persists with the height, whence it follows that $B=B_r \approx 1/R^2$ ($B_r$ is the field radial component). But open field lines of a large-scale field center only in coronal holes up to the $B \approx 1/R^2$ distance. Coronal holes, in turn, feature a super radial beam divergence of magnetic tubes leaving them. This implies that the sectional area of such magnetic tubes (as well as of its any portion), and, hence, the magnetic field value varies with height faster, than $1/R^2$. In all the other part of the corona, including its quiet part, the large-scale magnetic field comprises closed field lines (except for rare occurrences, when there are thin tubes with an open field in an active region). In the region of the magnetic field closed configuration, the field varies with height by the law different from the $B \approx 1/R^2$ dependence. One may show that, for some field configurations, the field value will vary with height more slowly, than $1/R^2$. In this case, the real values of the Alfv\'en velocity at each height will be greater, than those obtained by \citep{Mann1999}, because the plasma density dependence on the height varies insignificantly.

We used another approach to find the $V_A(R)$ dependence in the solar corona at $R>5R_s$. This dependence was calculated by using representative $B_r$ and $n_p$ (proton density) values in the Earth orbit and the known regularities of the variation in these parameters in the solar corona and in the interplanetary space. Because the CME axis may appear both in the fast solar wind from a coronal hole and in the slow wind in the coronal streamers forming the streamer belt \citep{Schwenn1990} or the streamer chains \citep{Eselevich1999,Wang2007}, we calculated separately the $V_A(R)$ dependences in the fast and slow wind, respectively. By using the OMNIWeb Plus site data (\url{http://omniweb.gsfc.nasa.gov}), one can realize that the $B_r = (4-8) \times 10^5$ G, $n_p = (2-7)$ cm$^{-3}$ are adequately representative values in the Earth orbit fast solar wind. We note again that these estimates relate to the Earth orbit solar wind fluxes originated in coronal holes on the Sun. In individual fast wind fluxes, the $B_r$ and $n_p$ values may deviate from the above, but insignificantly.

To estimate the $B_r$ and $n_p$ values in the Earth orbit slow wind, one should take into account the following. In most cases, the values of the magnetic field and of the proton density in the Earth orbit slow wind reflect the coupling between the fast and slow winds that exists in the interplanetary space. It is practically impossible to take into account the contribution of this coupling to the $B_r$ and, $n_p$ values at different distances from the Sun. To determine the values of the magnetic field and of the proton density at the distances in the corona that we are interested in $R=(5-25)R_s$ from the data on the Earth orbit magnetic field and plasma, we used the Earth orbit field and proton density values in the horizontal (almost parallel to the solar equator planes) segments of the streamer belt from \citep{Fainshtein1991}. In this case, there is practically no coupling between the fast and slow flows of the solar wind, and one may consider that we deal with the "true" parameters for the Earth orbit slow solar wind. The range of the $B_r$ and, $n_p$ representative values in the Earth orbit was $(3-8)\times10^5$ G and $(7 - 20)$ cm$^{-3}$ in this case.

We reveal now, by which law $B_r$ and, $n_p$ varies with distance. From the magnetic flux conservation, it follows that $B_r \approx 1/R^2$ from the Earth orbit to $R\approx(2.5-4)R_s$. For large distances $(R>2~AU)$, on average, this is shown to be, in fact, true from different satellite data \citep{Schwenn1990}. From the solar eclipse data, in a coronal hole, $n_e\approx1/R^2$ within $R\approx(4-8)R_s$. Here, $n_e$ is the electron density. Disregarding the difference between $n_p$ and $n_e$ in the coronal plasma, we regard $n_p\approx1/R^2$ at these distances. At $R=(8-30)R_s$, the $n_p\approx1/R2$ dependence in the fast solar wind from coronal holes may be also regarded true. This follows from the mass conservation equation (that, in the simplest case, can be written as $m_pn_pVR^2=const$), provided $V\approx const$ at these distances. Here, $m_p$ is the proton mass, $V$ is the fast solar wind velocity. In the slow wind, we assumed the dependence of $n_p(R) \approx n_e(R) = 3.3 \times 10^5R^{−2} + 4.1 \times 10^6R^{−4} + 8.0 \times 10^7R^{−6}$, сm$^{−3}$ \citep{Leblanc1998}. This dependence was used in a number of studies to determine the Alfv\'en velocity (see, for example, \citealp{Gopalswamy2011}). The coefficients in bold changed so that, in the Earth orbit, the $n_p$ density calculated by this relation possessed the above minimal or maximal value $(7-20)$ cm$^{-3}$ in the Earth orbit slow wind.

Figure 2(a) shows the calculated $V_A(R)$ dependences in the fast and slow solar winds at $R\approx(5-25)R_s$ considering possible variations in $B_r$ and $n_p$ in the Earth orbit for two types of solar wind fluxes. For comparison, the $V_A(R)$ model dependence from \citep{Mann1999} is also provided there. Note that the Alfv\'en velocity depends on the full value of the magnetic field, $B$. In the interplanetary space, there are three field components. Of the latter, the one perpendicular to $B_r$ (that also forms the Archimedean spiral) may be comparable with the $B_r$ far from the Sun surface. But one may show that, at $R\approx(5-25)R_s$, the difference between B and $B_r$ is relatively insignificant and is within the error satisfying us. Therefore, when building the $V_A(R)$ dependence in Figure 2, we assumed that $B\approx B_r$. In the slow wind, the $V_A(R)$ dependence obtained by \citealp{Mann1999} agrees most notably with the dependence that we obtained for the minimal values of $B_r$ and $n_p$ at the horizontal segment of the slow wind, starting with $R\approx7R_s$. For the maximal values of $B_r$ and np, the Alfv\'en velocity values obtained through the two techniques become close only at $R\approx(25-30)R_s$.

\begin{figure}[!ht] 
\centerline{\includegraphics[width=1.\textwidth]{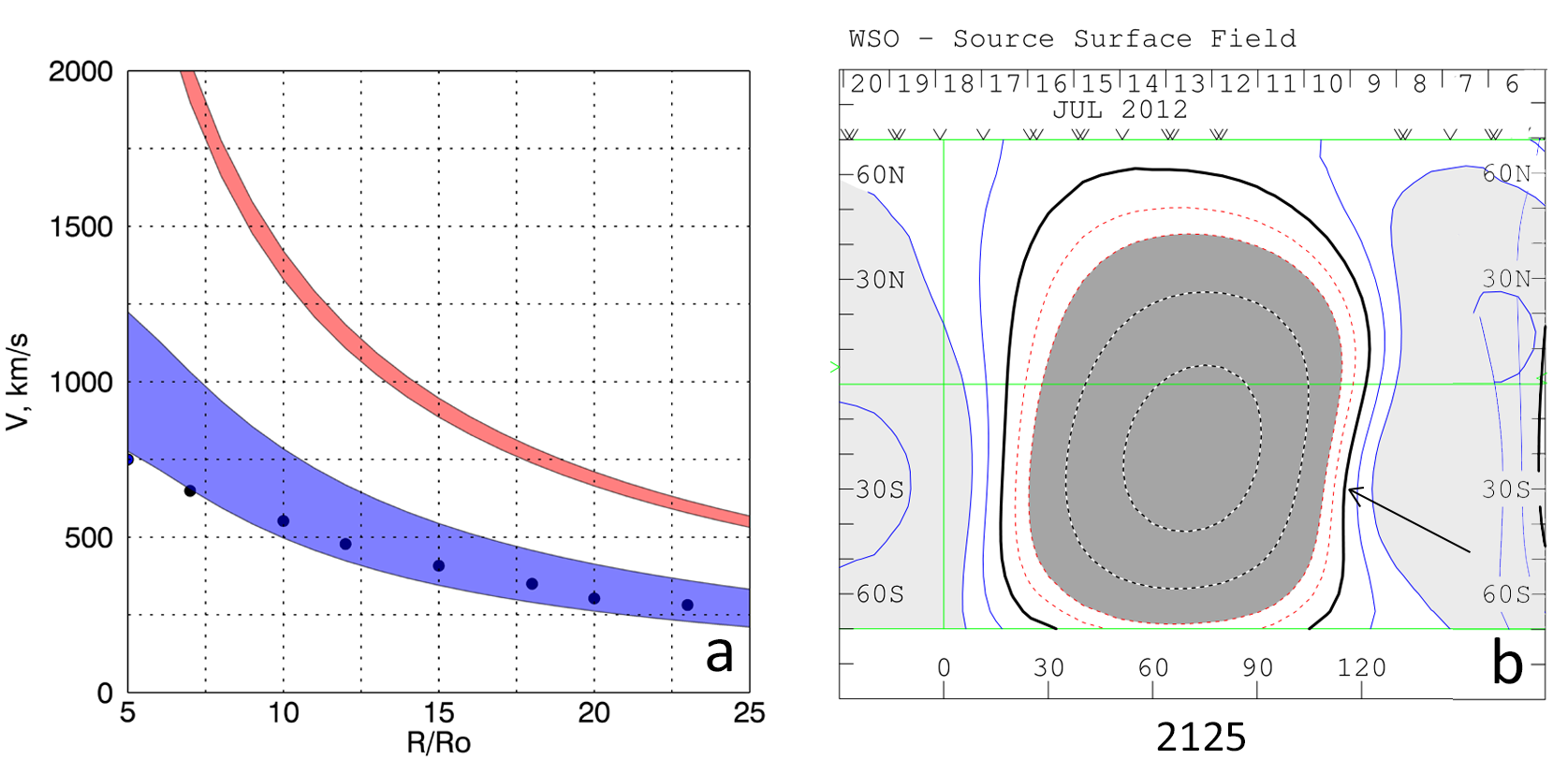}}
\caption{(a) Distance dependence of the Alfv\'en velocity, $V_A(R)$, in the fast (top band) and in the slow (lower band) solar wind. The $B_r$ and $n_p$ values in the Earth orbit at the $V_A(R)$ upper boundary of the detected $V_A(R)$ band are $B_r = 4 \times 10^5$ G, $n_p$ = 2 cm$^{-3}$ in the fast wind; at the lower boundary, $B_r = 8\times10^5$ G, $n_p$ = 7 cm$^{-3}$. In the slow wind, at the upper boundary of the $V_A(R)$ band, $B_r = 3\times10^5$ G, $n_p$ = 7 cm$^{-3}$; at the lower boundary, $B_r = 8\times10^5$ G, $n_p$ = 20 cm$^{-3}$. Dots show the $V_A(R)$ values at several distances. These $V_A(R)$ values were taken from the $V_A(R)$ dependence in \citep{Mann1999}. (b) Synoptic chart (CR 2125) of the magnetic field on the source surface with the $R=2.5R_s$ radius obtained from the field calculations in potential approximation (\url{http://wso.stanford.edu/synsourcel.html}). The thick black line shows the magnetic field polarity inversion line on the source surface (neutral line). The arrow indicates the neutral line segment that will appear over the Sun western limb on 2012 July 17.}%\label{fig:?}
\end{figure}  

In general case, the CME body different sites with relatively large angular size move both in the fast and in the slow solar wind. In the addressed event, the CME body paraxial part (extended along the latitude) appeared to be moving in the slow wind. The ground for this is the conclusion that, while observing the CME, over the western limb of the Sun there is a latitudinally extended segment of the magnetic field neutral line on the source surface in the magnetic field calculations in potential approximation, Figure 2(b). This figure shows the synoptic chart, on which provided are the latitude-dependent field radial component values on the central meridian (CM). This chart was obtained from the coronal magnetic field calculations in potential approximation at the Wilcox Solar Observatory (WSO) (\url{http://wso.stanford.edu/synsourcel.html}). 

The analyzed CME was recorded over the western limb. To find, which site of the Figure 2(b) synoptic chart was on 2012 July 17 over the western limb, it is enough to look at this synoptic chart site 6.8 days (a quarter of the solar rotation) before 2012 July 17. This site was at the central meridian approximately on 2012 July 10. On that day, the segment of the neutral line (dividing the magnetic field regions with opposite polarities) extended along the latitude (arrow) appeared to be in the central meridian surroundings. The heliospheric current sheet is known to correspond to the neutral line in the belt of coronal streamers, and the slow wind propagates there \citep{Schwenn1990}. This implies that a considerable part of the CME body moves in the streamer belt, where the slow solar wind propagates.

For this reason, we used the slow wind velocity as a solar wind velocity in the expected $V_b>V_A + V_{SW}$ condition of the shock formation. As the $V_{SW}(R)$ dependence, we used the expression for the slow wind velocity obtained in \citep{Sheeley1997} for $R>4.5R_s: V^2_{SW} (R) = 1.75 \times 10^5 (1 - exp (- (R/R_s - 4.5)/15.2))$. This expression was already used earlier to determine the Alfv\'en velocity within the LASCO coronograph fields-of-view \citep{Gopalswamy2011}.

\section{Results}
     \label{S-Results} 

Fig. 3(b) enables to estimate the site, where the possible $V_b > V_A + V_{SW}$ condition of the CME-related shock generation within the LASCO C3 is met. In this figure, the strip shows the $V_A + V_{SW}$ distance dependence in the slow wind for different $B_r$ and $n_p$ values in the Earth orbit, as well as the CME body velocity distance dependence $V_b(R)$ approximately along the CME body axis. Apparently, the CME body velocity crosses the total of the Alfv\'en velocity and the slow solar wind velocity within $(8.1 - 12.3) R_s$ and in period of time from 15:52 UT to 16:50 UT. Note that we took into account the mean squared scatter of the slow wind velocity which is close to 75-85 km/s.

\begin{figure}[!ht] 
\centerline{\includegraphics[width=1.\textwidth]{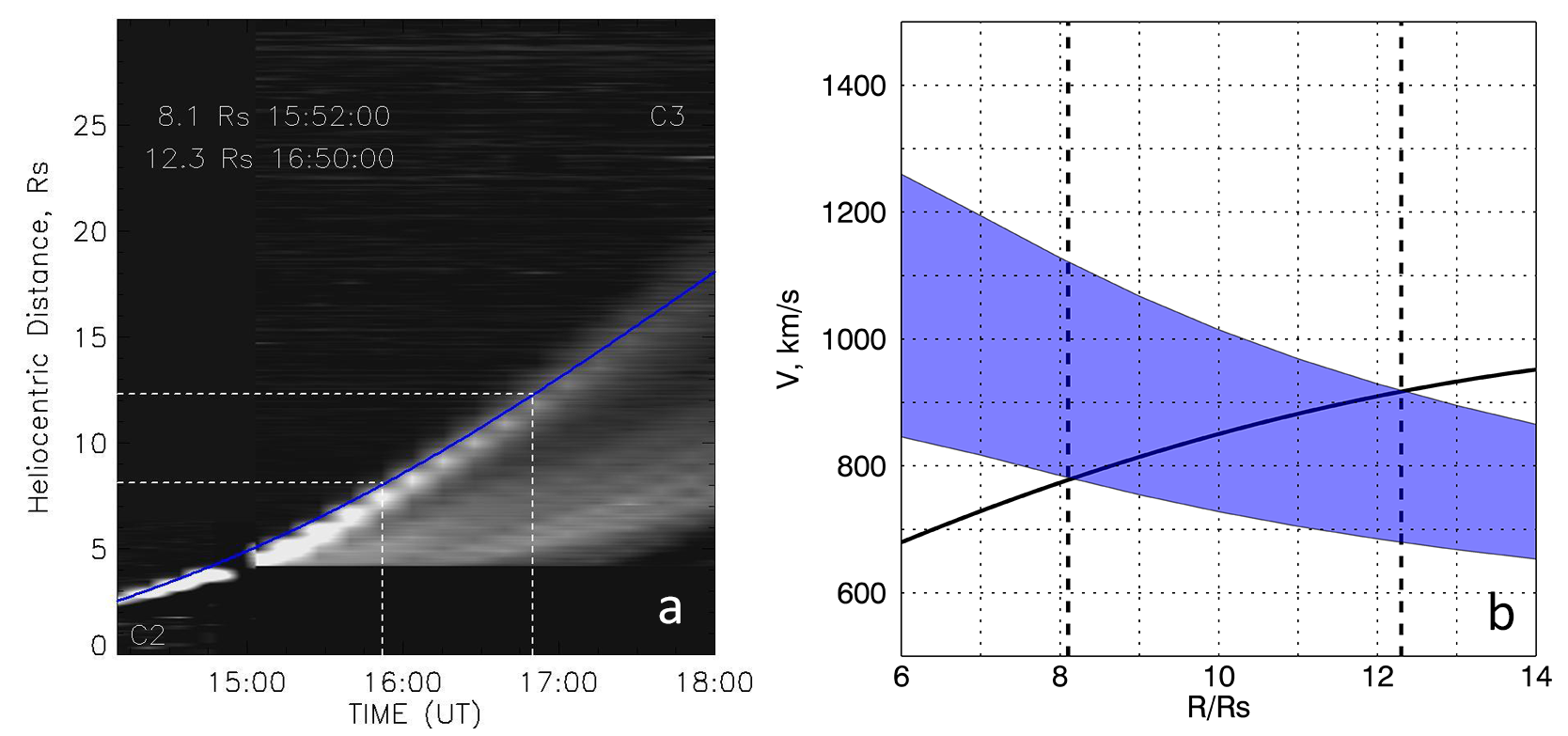}}
\caption{(a) The stack-plot obtained from LASCO C2 and C3 images. The blue solid line marks the time dependence of the CME body $R_b(t)$; (b) The strip shows the distance dependence of the total between the Alfv\'en velocity $V_A(R)$ and the slow solar wind velocity $V_{SW}(R)$. The black solid line marks the distance dependence of the CME body velocity $V_b(R)$. }%\label{fig:?}
\end{figure}  

Hence, it follows that, to estimate the implementation of the shock generation condition, one should compare the relation between the CME body velocity $V_b$ and the total velocity $V_A(R) + V_{SW}(R)$, where $V_{SW}(R)$ hereafter is the slow wind velocity at the R distance from the solar disk center. We assume that the shock will be generated in the region of the latitudinally-extended segment of the slow solar wind, when the condition $V_b>V_A + V_{SW}$ is met.

From Fig. 4, it follows that at t=16:40 UT, the diffuse region brightness decreases sufficienty smoothly to the background values ahead of this region boundary. The brightness jump formed in the next moment of time at $R>11.5R_s$ with width equal to the spatial resolution of the LASCO C3 coronagraph ($0.125R_s$). After that, the formed brightness jump moves forward without a width change. We believe that this brightness jump is a collisionless shock \citep{Artsimovich1979}, because this brightness jump width is manyfold less, than the mean free path of the coronal plasma charged particles. After the shock formation, the diffuse region ahead of the CME body converts into a «sheath» between the shock and the CME body, or (which is the same) into the shock-compressed plasma region behind the shock front. 
 
\begin{figure}[!ht] 
\centerline{\includegraphics[trim=0.0cm 0cm 0.5cm 12cm, width=1.0\textwidth]{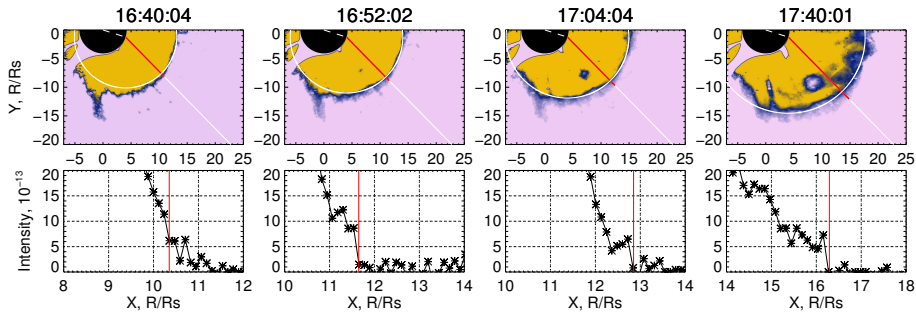}}
\caption{ Brightness distribution along the beam. The red vertical line shows the brightness jump which is a collisionless shock.}%\label{fig:?}
\end{figure}

Below, we attempt to reveal, what mechanism for the CME-related shock formation is, and what the features of the variations in the related shock characteristics are during the shock formation and during the initial stage of its existence.

Note that the origin of the CME and of the related shock was preceded by the emergence of a flux-rope well observed in the SDO/AIA 193 \AA (Fig. 5). Like seen from Fig. 5, the fux-rope motion from observations in that channel features slight velocities and accelerations. This means that the shock generation probability at these heights by the moving flux-rope, according to the mechanism proposed in \citep{Grechnev2011} is insignificant. Our analysis showed that, within the LASCO C2 FOV, the CME core that (like we assume) is the flux-rope observed in 193 \AA may reach the velocity of $\approx 1150~km/s$ and the accelerations of $0.96~km/s^2$. In this case, one may expect a shock generation by a sharp effect of an eruptive core on the higher magneto-plasma structures. But we found no shock origin in such a manner.

\begin{figure}[!ht] 
\centerline{\includegraphics[trim=0.0cm 0cm 0.5cm 12cm, width=1.0\textwidth]{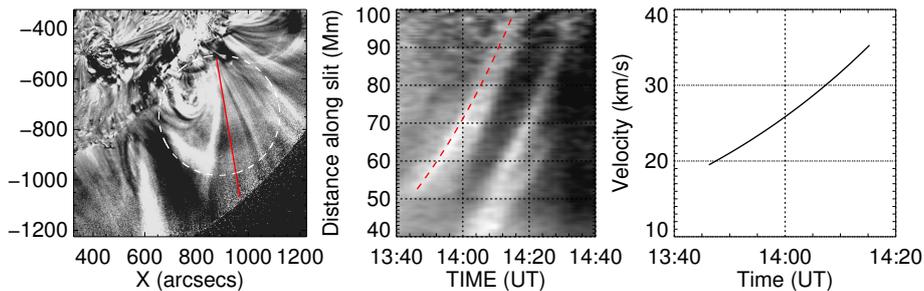}}
\caption{Panel (a) is the difference image of the corona in SDO/AIA 193 \AA, where one can see an eruptive rope, whose outer boundary is encircled by the dot line. In addition, there is also a straight line, along which the brightness was scanned; panel (b) is the time dependence of the rope velocity along the red beam; panel (c) is the time dependence of the rope acceleration. }%\label{fig:?}
\end{figure} 

Apparently, in the addressed event, we deal with a simultaneous implementation of two mechanisms for a shock generation. In the Introduction, we already noted that it is possible taking the CME body motion features into account (the CME body moves translationally and, simultaneously, it expands, which reflects in its angular size weak change). Here, we present the observational results that support this conclusion. First of all, we note again that, starting with $R \approx(8.1-12.3)R_s$ from the solar disk center, the CME body moves relative to the ambient slow solar wind at the longitudinal velocity exceeding the Alfv\'en velocity. This implies that the coronal plasma surrounding the CME body flows around it at the Superalfv\'en velocity in CME body axis surroundings, and, thereby, the necessary condition of a bow shock generation is met. Figure 6(c) supports this assumption. From this figure, it follows that the distance between the shock and the CME body increases with time, as the CME body transversal size grows. The larger the radius of the boundary surface on the CME body axis, the father the bow shock from the body flown around by gas (plasma). We also compared the radial velocities of the CME body ($V_b$) and of the related shock ($V_{SH}$) (Figure 6(b)). Apparently, that with time(eventually) the shock velocity $V_{SH}$ increases eventually faster, than the CME body velocity $V_b$. At large distances, the $V_{SH}-V_b$ difference exceeds 250 km/s (Figure 6(d)). 

\begin{figure}[!ht] 
\centerline{\includegraphics[width=1.\textwidth]{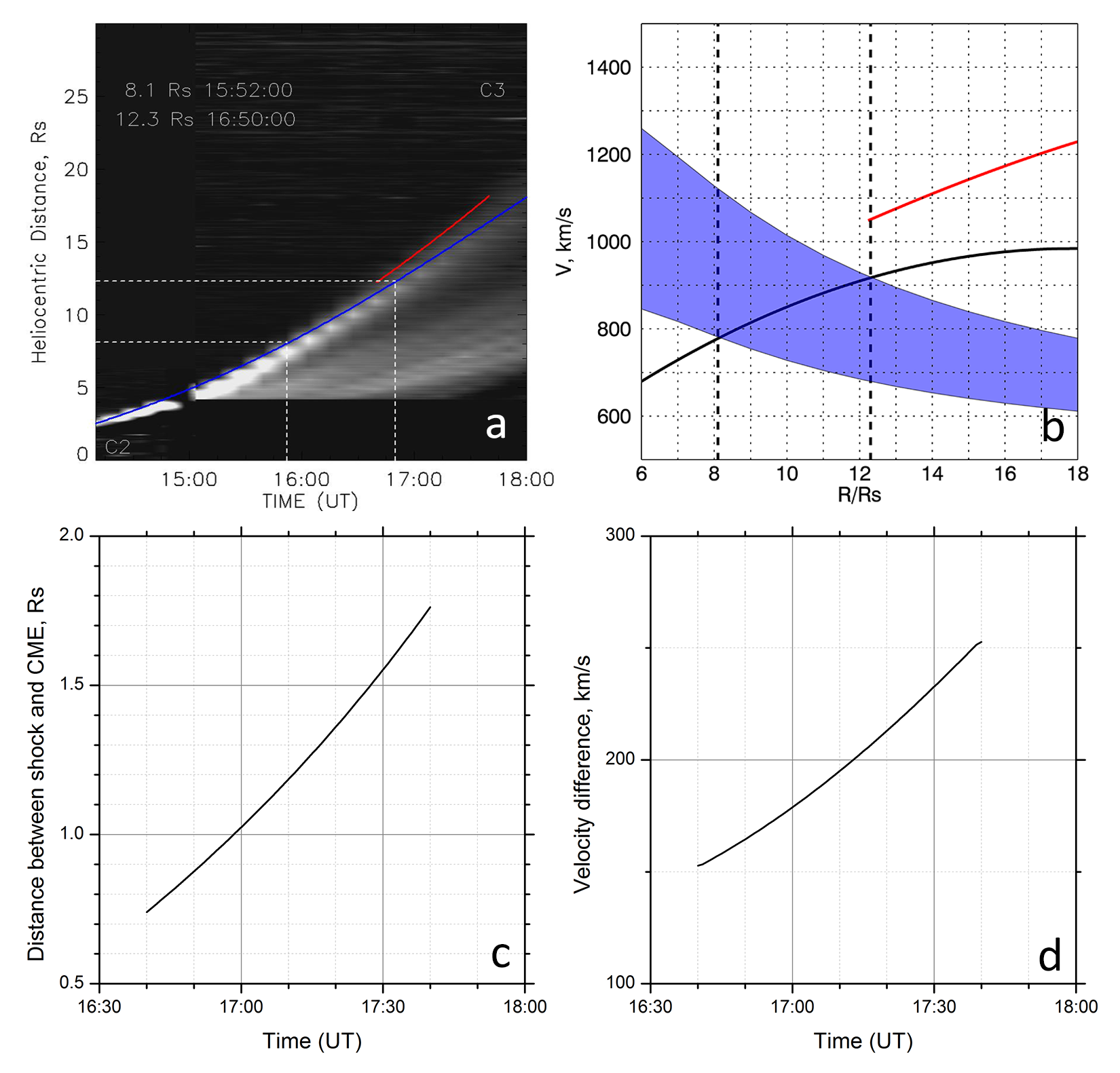}}
\caption{(a,b) The same as Figure 3(a,b) but with shock postion and velocity reprecented by red solid curve; (c) Time dependence of the distance between the shock and the CME body; (d) $V_{SH} - V_b$ }%\label{fig:?}
\end{figure}  

 Also, we attempted to estimate the CME body translational motion velocity ($V_t$). The difference between $V_b$ and $V_t$ is caused by that $V_b$ is a resultant velocity of the CME body complex motion in the radial direction. This motion includes the CME body translational motion and its expansion in the radial direction. Herewith, $V_t$ is the velocity of only translational motion. 

To estimate the CME body translational velocity, we used the ellipse fitting routine (Figure 7(a)) which allow us to extract longitudinal (semi-minor axis) and lateral (semi-major axis) expansion of the CME. Then we subtract this longitudinal expansion from CME velocity and obtained the CME body translational velocity (Figure 7 (c) solid line and crosses). Another way to find $V_t$ is to obtain time dependence of ellipse center propagation velocity (circles and dashed line).  One can see that the CME body translational velocity appeared relatively small, and it exceeds the total velocity $V_a+V_{SW}$ nowhere. This implies that, as long as the shock originates, because the ambient plasma flows around the CME body at the superAlfv\'en velocity relative to the solar wind, then, unlike the gas flow around a solid body with invariable size (bullet, airplane, rocket), here, the ambient plasma flow around the CME originates due to both the translational motion and the CME body expansion.

\begin{figure}[!ht] 
\centerline{\includegraphics[width=1.\textwidth]{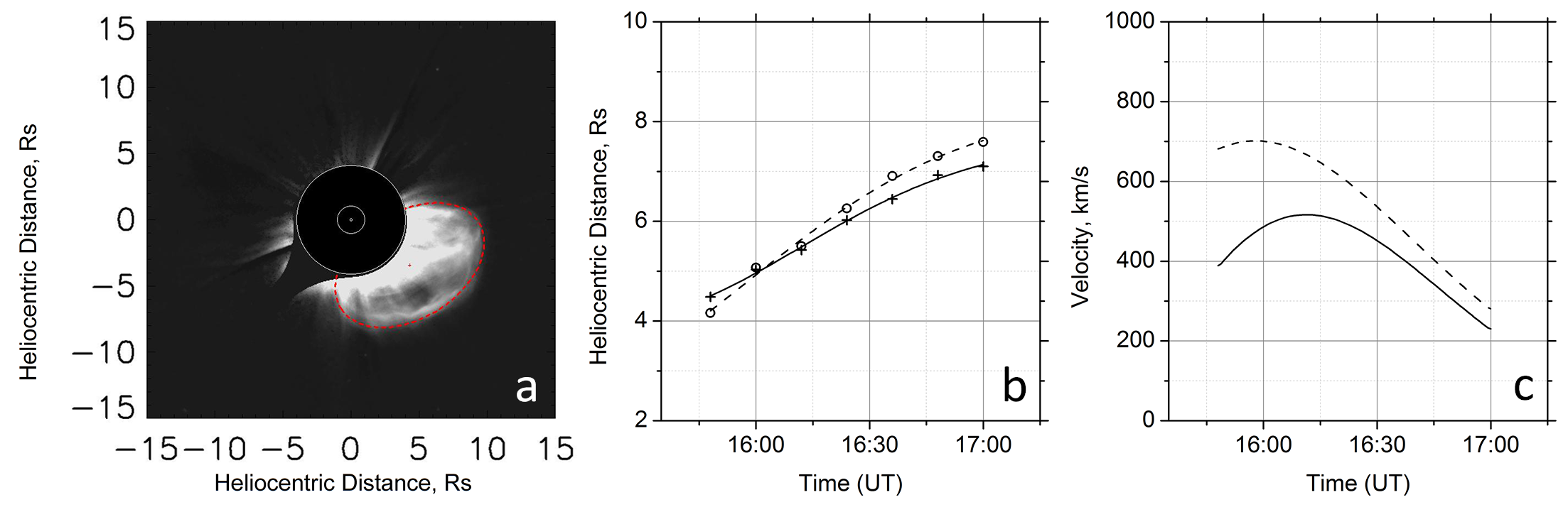}}
\caption{a) the CME front fitted by red ellipse. body image (turned relative to the CME body true position so that the CME axis became vertical to convenienедн determine the CME body parameters) with the segments from the CME body axis to the boundaries of the CME body most distant side features; c) solid line is the CME body translational velocity equal to the difference between the CME front velocity and its longitudinal expansion rate. Dashed line is the ellipse center propagating rate.}%\label{fig:?}
\end{figure}  

From Fig. 7(c), it follows that the CME body translational velocity is relatively insignificant, and, as long as the CME did not expand at a sufficiently high velocity, the bow shock ahead of the moving mass ejection body would not originate.

And, finally, we estimated the value of the Alfv\'en Mach number $M_a = (V_{SH} - V_{SW})/V_a$ for the shock after its origin. The Ma value is established to grow at the motion initial stage according to the Vsh growth and the $V_a$ decrease from $M_a$ = 1.2 to $M_a$ = 1.95. Later, it slightly decreases, and further varies insignificantly. The mean Alfv\'en Mach number is 1.7.

\section{Discussion and conclusions}
     \label{S-Conclusion} 

Ahead of fast CMEs within the LASCO С2 and/or С3 fields-of-view, shocks are observed \citep{Vourlidas2012}. At the same time, physical mechanisms for the CME-related shock generation, as well as details of their formation, have not been practically studied. There exists a conclusion that the CME-related shocks are bow ones \citep{Vourlidas2003}. In some papers, such shocks are considered piston \citep{Chen2011}. But the mechanisms for and the conditions of the bow and piston shock generation differ radically. A piston shock originates due to the compression of the surrounding medium by a piston moving translationally, and owing to the subsequent non-linear gas (plasma) disturbance evolution. Herewith, a shock may be generated, including, as long as the piston velocity is lower, than the speed of sound in gas (in our case, it is less, than $V_A + V_{SW}$). The shock formation site and the shock amplitude depend on the relation between the piston velocity and the speed of sound in gas (or between $V_b$ and $V_A$ + $V_{SW}$). But the shock velocity ($V_{SH}$) will always be higher, than the speed of sound (for our event, $V_{SH}> V_A + V_{SW}$) \citep{Sedov,Zeldovich}. 

A bow shock originates only provided that the upstream gas(plasma) flow velocity on a resting body or the body velocity relative to gas is more, than the speed of sound in gas (or, for the body moving in the coronal plasma, $V_b$ and $V_A$ + $V_{SW}$). As long as the difference between the body velocity (or the velocity of the gas upstreaming the body) and the speed of sound is insignificant, a shock originates far from the body. As the difference between the velocity of the body flown around by gas and the speed of sound increases, the shock nears the body. Provided the body has a sharp tip, the shock will practically touch this tip. As long as the boundary of the streamlined body features a large curvature radius, the shock moves away from the body along the axis of the latter. Herewith, the more the body boundary curvature radius, the farther the shock moves away from the body \citep{Landau_EN}.

The CME body motion includes two components: a translational motion in the radial direction and a simultaneous expansion. The transversal expansion manifests itself as a weak variation in the angular size of many CMEs as they move. As a result, one may expect that, at such a CME body motion, both mechanisms for the CME-related shock generation participate, but the process of the shock origin, in this case, and its characteristics at various stages of their formation and their subsequent motion remain obscure.

In this study, we investigated the CME-related shock origin within the LASCO C3 FOV, and the features of the shock motion at the initial stage after its origin. The LASCO C2 and C3 observed this CME on 2012 July 17. The features of this CME are that its frontal structure originates at a relatively great height (it was not observed within the AIA telescope fields-of-view) and that its velocity accelerates slowly with time (distance). Apparently, for the first time, we detected the CME-related origin within the LASCO С3 FOV in the form of a sharp brightness jump. The shock is generated at the distance exceeding the one, at which the CME body radial velocity surpasses the sum of the Alfv\'en velocity and the slow solar wind velocity ($V_b$ and $V_A$ + $V_{SW}$). This enables to assume that this is a bow shock, i.e., mainly, its emergence is caused by the surrounding coronal plasma flowing around the CME body at a "super-Alfv\'en" velocity relative to the solar wind. This assumption supports another observation: the distance between the shock and the CME body increases with time (with distance) as the CME body grows. We note that, in this case, the CME body radial velocity is the total of the translational velocity and the CME body expansion rate along the motion direction. Our analysis showed (see below) that the CME body translational velocity is less, than $V_A + V_{SW}$. This implies that the flow around the СME body only at its translational motion velocity does not provide the condition necessary for the bow shock generation.

The LASCO insufficient temporal resolution does not enable to see the process of the shock formation in all the details. Some data testify to an eventual steeping of the coronal plasma of the disturbed area ahead of the CME body, as a result of which the shock emerges. Thus, the CME-related shock generation is accompanied by meeting the conditions necessary for the formation of both bow and piston shocks. But the physical pattern of forming the shock moving translationally and, simultaneously, expanding has remained obscure. One should study a large sampling of events, and, desirably, from the data with a higher temporal resolution of coronagraphs.  

We arrive at the conclusion that the completely formed shock is collisionless, because its observed brightness front width is close to the coronagraph spatial resolution that is manyfold less that the mean free path of the plasma charged particles between their collisions (e.g., electrons with protons or protons with protons).

We compared the shock and the CME body velocities at different distances. Both velocities grow with distance, but the shock velocity grows faster, than the CME body velocity. We estimated the CME body translational velocity depending on the distance. It varies within 130 - 230 km/s, and, at all the distances, it appears less than the maximal rate of the CME body transversal expansion.

We obtained the alfv\'en Mach number at several distances. The mean value of such a Mach number for the mean alfv\'en velocity at each distance was 1.5, the maximal value being 2.

%%%%%%%%%%%%%%%%%%%%%%%%%%%%%%%%%%%%%%%%%%%%%%%%%%%%%%%%%%%%%%%%%%%%%%%%%%%
%% Appendix
%
% \appendix   

%%%%%%%%%%%%%%%%%%%%%%%%%%%%%%%%%%%%%%%%%%%%%%%%%%%%%%%%%%%%%%%%%%%%%%%%%%%
%% Acknowledgements
%
 \begin{acks}
The authors thank the LASCO, SDO/AIA, GOES WSO teams for a possibility to use their data free; also, the authors are grateful to A.M. Uralov and V.V. Grechnev for useful discussions and method assistance. The study was done with a support from grants by the Russian Foundation for Basic Research Nos. 15-02-01077-а and 16-32-00315. 
 \end{acks}

%%% %%%%%%%%%%%%%%%%%%%%%%%%%%%%%%%%%%%%%%%%%%%%%%%%%%%%%%%%%%%
%% Bibliography
%
% Using BibTeX
%
 \bibliographystyle{spr-mp-sola}
 \bibliography{biblio2}

\end{article} 
\end{document}